\documentclass[12pt]{article}

\usepackage{latexsym}
\usepackage{amssymb}
\usepackage{amsfonts}

\usepackage{xcolor}
\usepackage{graphicx} 
\definecolor{brownn}{cmyk}{0,1,1,0.5}
\definecolor{bluen}{rgb}{.1 ,0, .8}

\RequirePackage[T1]{fontenc}
\RequirePackage{times}
\usepackage[hypertexnames=false,
colorlinks,linkcolor=bluen,citecolor=brownn,urlcolor=brownn]{hyperref}
\graphicspath{{images/}}

\def\rnc#1{\renewcommand{#1}}

\renewcommand{\b}{\beta}

\newcommand{\ep}{\epsilon}

\renewcommand{\k}{\kappa}

\newcommand{\La}{\Lambda}

\newcommand{\non}{\nonumber\\}
\newcommand\nn{\nonumber}

\newcommand{\p}{\partial}

\newcommand{\x}{\times}

\newcommand{\beq}[1][]{\begin{equation}\label{#1}}
\newcommand{\eeq}{\end{equation}}

\rnc{\-}{\item}

\renewcommand{\to}{\ensuremath{\rightarrow\;}}

\newcommand{\oto}{\leftrightarrow}

\usepackage{mathcomp}
\usepackage{amsmath}

\newcommand{\bt}{b^-}

\newcommand{\rhot}{{\rho_t}}

\newcommand{\av}[1]{\langle {p{^#1}}\rangle}

\renewcommand{\sl}[1]{\frac{#1}{2\pi}}
\newcommand{\isl}[1]{\frac{2\pi}{#1}}
\newcommand{\dr}[1]{#1^{\rm dr}}
\newcommand{\mi}[1]{{\hat{#1}}}

\newcommand{\be}{\mathfrak{b}}
\newcommand{\bet}{\mathfrak{b}^-}
\newcommand{\bep}{\mathfrak{b}^+}

\newcommand{\mf}{\mathfrak{m}}

\newcommand{\rhs}{r.h.s.}
\newcommand{\cI}{{s}}
\newcommand{\ads}{AdS$_3$}
\newcommand{\adsst}{\ads$\x$S$^3\x$T$^4$ }
\newcommand{\qn}{\Psi}

\newcommand{\nat}{Nature}
\newcommand{\prl}{Physical Review Letters}

\begin{document}
\begin{center}{\Large \textbf{
Deforming integrable models of AdS3 strings
}}\end{center}

\begin{center}
J.~Pawe{\l}czyk \textsuperscript{*}
\end{center}

\begin{center}
  Faculty of Physics, University of Warsaw,  Pasteura 5, 02-093 Warsaw, Poland
\\
*  jacek.pawelczyk@fuw.edu.pl
\end{center}

\begin{center}
\today
\end{center}


\section*{Abstract}
{
We discuss an integrable model of string on AdS$_3$xS$^3$xT$^4$ in a thermodynamical bath.
We show that scattering of the excitations above equilibrium states has some novel features. 
Thermodynamics points to interesting deformations of the original  model for 
which we  discuss finite size effect through mirror TBA equations.  
}
\newpage
\tableofcontents

\section{Introduction}
The AdS/CFT correspondence \cite{1999IJTP...38.1113M},\cite{1998PhLB..428..105G,1998AdTMP...2..253W}
(see e.g. the first review \cite{2000PhR...323..183A}) sparked incredible research activities ranging from basics of string theory to condensed matter physics \cite{2009CQGra..26v4002H,2009arXiv0909.0518M,2012sta..conf..707I,Zaanen:2015oix}.  It appeared that the some string models, including  AdS$_5\x$S$_5$, can be realized as an integrable system \cite{2003JHEP...03..013M,2004PhRvD..69d6002B} (for the review see \cite{2012LMaPh..99....3B}). Unfortunately the proposed models have very non-trivial scattering matrices. Reformulation of the theory into so-called Y-system \cite{2012LMaPh..99..321G} simplified many calculations considerably.
Developments for \ads{} strings showed similar difficulties \cite{Babichenko:2009dk,Cagnazzo:2012se,Lloyd:2014bsa,Borsato:2015mma}.
Recent works on QSC provide a more complete understanding of these holographic models \cite{2021arXiv210905500C,2021arXiv210906164E}.
Besides these successes intricate difficulties discouraged numerous investigators \cite{Bajnok}.

On the other hand  it was shown that ordinary bosonic string 
with one dimension compactified on a circle can be represented as an integrable model with very simple diagonal scattering matrix \cite{2012JHEP...09..133D}. 
Few year later sightly modified construction yielded spectrum of strings on 
\adsst \cite{Baggio_2018,Dei_2018} and on \ads$\x$S$^3\x$S$^3\x$S$^1$ \cite{2019JHEP...02..072D} with pure NSNS flux. Since then the subject has an interesting development,
see e.g. recent works \cite{2020arXiv201002782S,2021PhRvD.103j6024R}.

In this letter we dwell upon \adsst model of \cite{Baggio_2018,Dei_2018}. Specifically we analyze its thermodynamical behaviour. It is known that integrable models have stable excitations above thermodynamic equilibrium but their energies, momenta and the interaction are subject to finite renormalization called "dressing". 
We calculate dressed scattering matrix and dressed spectrum of excitations. 
Inspired by these results we introduce a new model with general chemical potentials and  modified scattering phase for which we derive formulae for its finite size spectrum.

The paper is organized as follows: we begin recalling basic facts from
\cite{Baggio_2018,Dei_2018}. In Sec.\ref{sec:TBA} we put the system in a thermodynamic bath  parameterized by temperature and chemical potentials which renormalizes states and interactions.
The procedure can be considered as deformations of the original theory.
We derive  dressed scattering matrix and  energies of elementary excitations. These results will be our starting point for the next
section, \ref{finite}, where we define the new model and calculate its finite size  spectrum. Finally in Sec.\ref{sec:cp} we introduce CP symmetry.

\section{The \adsst Model}\label{sec:init}
In papers \cite{Baggio_2018,Dei_2018} an interesting integrable model of string theory on \adsst was introduced. It was shown there and in subsequent papers that the the model reproduces spectrum of the  theory. Here we present its main ingredients which are the set of quasi-particles (see the table), their dispersion relations and the scattering phases defining integrable interactions.
\begin{table}
\centering
\renewcommand{\arraystretch}{1.2}
\begin{tabular}{|c|c|c|c|c|c|c|c|c|c|}
\hline
& $Y(p)$&$\bar Y(p)$ &$Z(p)$ & $\bar Z(p)$ &$T^{a\dot{a}}(p)$ &$\eta^a(p)$ &$\bar \eta^a(p)$ &$\chi^{\dot{a}}(p)$ & 
$\bar{\chi}^{\dot{a}}(p)$ \\ 
\hline 
$\mu$ & $1$  &$-1$ &$1$ &$-1$ &$0$ &$1$ &$-1$&$0$&$0$      \\
\hline
\end{tabular}\label{tab:particles}
\caption{%
The particle spectrum of pure-NS-NS \adsst strings integrable model. 
Different quasi-particles  and anti-quasi-particles (entries with bars) are indexed with capital letters $A,B,C$ and will be called flavour.  Quasi-particles $Y,Z,T^{a\dot a},...$ are bosons for which $\psi_A=0$ while $\eta^a, \chi^a,..$ are fermions ($a,\dot a=1,2$) for which $\psi_A=i\pi$.}
\end{table}
 Dispersion relation of the modes  depends on the flavour $A$ through value of the chemical potential $\mu_A$
\begin{align}
\label{energy-0}
H_A^{(0)}(p)=\left|p+\mu_A\right|
\end{align}
Each flavour $A$ can be  chiral $p>-\mu_A$ or  anti-chiral $p<-\mu_A$.
"Zero-modes", which are neither chiral nor anti-chiral, have momenta $p=-\mu_A$ and  vanishing energy. We shall assume that they will not play any role hereafter. 

The scattering phase is diagonal and depends on flavours $A,B$ only through definition of chirality \footnote{Our momenta $p$ are rescaled compared to \cite{Dei_2018} by the  factor $\sl{k}$.}.
\begin{align}\label{scatt-0}
\Phi_{AB}^{(0)}(p,q)=\mp\isl k p\, q\ ,\qquad\left\{\begin{array}{ll}
-&\mbox{for }\;p>-\mu_A, q<-\mu_B\\
+&\mbox{for }\;p<-\mu_A,q>-\mu_B
\end{array}\right.
\end{align}
We shall denote with superscript $\pm$ quantities related to the chiral or anti-chiral quasi-particles e.g. $p_A^\pm$ are momenta $p>-\mu_A$ for $(+)$ and  $p<-\mu_A$ for $(-)$  so the scattering phase can be expressed as
$\Phi_{AB}^{(0)}(p,q)=-\isl k(p_A^+ q_B^- -p_A^- q_B^+)$.

The simple  product form of \eqref{scatt-0} leads to dramatic simplifications
of the TBA equations which turn to be transcendental equations on some unknown constants.
It is obvious that the scattering phase is not continues at $p=-\mu_A\neq 0$.
According to the assumption stated about "zero modes" we shall ignore terms resulting from differentiation of this discontinuity.

The spectrum of the model at finite but large string size $R$ is given by the Bethe equations (BE) which read:
\begin{align}
\label{BE-0}
p_{i}R+\sum_{j\neq i} \Phi_{ij}^{(0)}(p_{i}, p_{j})=2\pi n_{i}
\end{align}
We denote the i-th momenta of quasi-particles as $p_i$, the corresponding flavour as $A_i$ and the chirality scale  as $\mu_{A_i}$. The above equation naturally splits into chiral components. In similar manner to the positive/negative chirality momenta we split the integers $n_i$ of \eqref{BE-0} into $n_i^\pm$.
Hence \eqref{BE-0} can be rewritten as:
\begin{align}\label{BE-1}
p_{i}^+(R\,-&\isl k P^-)=2\pi n_i^+,\non
p_i^-(R\,+&\isl k P^+)=2\pi n_i^-
\end{align}
where $P^\pm=\sum_{j}  p_j^\pm$.
The equations were analyzed in \cite{Baggio_2018,Dei_2018} and it was shown that their solutions corresponds to string spectrum on \ads$\x$S$^3\x$T$^4$ with NSNS flux.
\subsection{Thermodynamics}\label{sec:TBA}
Integrable systems have interesting thermodynamical properties
\cite{Korepin:1993kvr, Mussardo:2010mgq}. Static behaviour are governed by TBA equations which yields  densities of quasi-particles at equilibrium.
Out-of-equilibrium phenomena can also by effectively analyzed using global or local quench protocols. One can also find the spectrum of excitations above the equilibrium. They appear to be stable and their scattering matrix is factorizable but their dispersion relation is modified through interaction with environment. The process in this context is called dressing. 
Summarizing we can say that  finite temperature theory is an integrable deformation of the zero temperature theory of dressed quasi-particles which interacts with dressed scattering matrix.
In this section we describe properties of dressed quantities for the model presented in the previous section. 

First of all  we
need to go to thermodynamical limit for BE of the previous section. This leads to so called BYE which total relates density of states $\rho_{tA}$ to density of quasi-particles 
$\rho_A$ for each flavour $A$.
\begin{equation}\label{bye-0}
\frac{1}{2\pi}+\sum_B\int \frac{dq}{2\pi}\ \phi_{AB}^{(0)}(p,q)\rho_B(q)
=\rhot_A(p)
\end{equation}
where  $\phi_{AB}^{(0)}(p,q)=\p_p\Phi_{AB}^{(0)}(p,q)$.
Explicitly:
\begin{align}\label{density-t}
\frac{1}{2\pi}-\isl k\av-&=\rhot^+_A,\qquad \frac{1}{2\pi}+\isl k\av+=\rhot^-_A
\end{align}
where,
\begin{equation}\label{pis}
\av-=\isl k\sum_B\int_{-\infty}^{-\mu_A}\sl{dq}\ q \,\rho_B^-(q),\quad 
\av+=\isl k\sum_B\int^{\infty}_{-\mu_A}\sl{dq}\ q \, \rho_B^+(q)
\end{equation}
are momenta carried by chiral and anti-chiral quasi-particles.
Thus total occupation numbers $\rhot_A^\pm$ are constants  dependent  on the chirality $(\pm)$ only.

 Equilibrium densities  $\rho_B(q)$ are  determined by TBA equations which are integral equations on
pseudo-energies $\ep_A(q)$. The latter determines equilibrium occupation numbers
$n_B(q)=({e^{\ep_B}-e^{\psi_B}})^{-1}$
and densities of quasi-particles which according to \eqref{density-t} are given by
$\rho_A(q)=n_A(q)\rhot$.
\begin{align}
\label{TBA-0}
\ep_A(q)&=w_A(q)-\int \frac{dp}{2\pi}\La_B(p)\ \phi_{BA}^{(0)}(p, q)
\end{align}
where\footnote{$\psi_A$'s are defined in the caption under Tab.\ref{tab:particles}.},
\begin{equation}
\La_B(p)=-e^{\psi_B}\log(1-e^{\psi_B}e^{-\ep_B(p)})
\end{equation}
Temperature $T=1/\b$ appears through
\begin{align}\label{pot}
w_A^+(q)&=\b(q+\mu_A+b^+_A),\ \quad w_A^-(q)=\b(-(q+\mu_A)+\bt_A)
\end{align}
In principle $w(q)$ could be arbitrary generalized potential as dictated by GGE \cite{2007PhRvL..98e0405R,2008Natur.452..854R} \footnote{See reviews \cite{2016RPPh...79e6001G,2016JSMTE..06.4007V} and also \cite{Hernandez-Chifflet:2019sua}.}, but here we limit our choice introducing
only the extra "chiral chemical potentials",
 $b^\pm_A$, which will be a free parameter in our considerations. The potentials are necessarily
non-zero as we shall see below.

We can solve \eqref{TBA-0} realizing that the  integral term on the rhs of \eqref{TBA-0} is proportional to momentum $q$ what implies that
\begin{align}
\label{e-dr}
\ep_A^+(q)=&w_A^+(q)-\,I^-q=(\b- I^-)(q+\mu_A)+\bep_A,&\quad q>-\mu_A
\non
\ep_A^-(q)=&w_A^-(q)+\, I^+q=-(\b- I^+)(q+\mu_A)+\bet_A,&\quad q<-\mu_A
\end{align}
where $\bep_A=\b b^+_A+\mu_A\, I^-,\ \bet_A=\b\bt_A-\mu_A\, I^+$ and
\begin{equation}
\label{Is}
I^\pm=\isl{k}\sum_A\int_{c^\pm}\!\sl{dq}\,\La_A(\ep^\pm_A)
\end{equation}
with $c^+=\{-\mu_A,\infty\},\ c^-=\{-\infty,-\mu_A\}$.
Hence \eqref{Is} are two  equations on $I^\pm$. 
These can be written as
\begin{align}\label{Is-eq}
I^+(\b-I^-)=& \isl{k}\cI(\bet),\quad
I^-(\b-I^+)=\isl{k}\cI(\be^+)
\end{align}
where the \rhs{} of above depends only on  $\be^\pm_A$. Explicitly:
\begin{align}
\label{int-1}
\cI(\be)&=\sum_A(-)\,e^{\psi_A}\!\!\int_0^\infty\sl{dq}\log(1-e^{\psi_A}e^{-q-\be_A}).
\end{align}

It is easy to find that  $0\leq s\leq 2\pi^2$ for non-negative $\be$'s.
The transcendental  equations \eqref{Is-eq} can be solved numerically while $\b$ and all $b^\pm_A$'s are fixed.  
However, notice that $I^\pm$ enter \eqref{int-1} through $\be^\pm_A$'s only i.e. in linear combination with chiral chemical potentials $b^\pm_A$'s  which are free parameters of the model. 
Thus we can treat $\be^\pm_A$'s as free parameters solving for \eqref{Is-eq} by fixing two independent $s(\be^\pm)$. We see that $b^\pm_A\geq0$ is required for reality of each term in \eqref{int-1}. 

\subsection{Dressing}

Integrable models have stable excitations above the thermal equilibrium  \cite{Korepin:1993kvr} (Sec.8). The single excitation modes have dispersion relation given by $\ep^\pm$ of the previous section - these we call dressed energies recalling that the "dressing" is thermodynamical effect. It appears that not only energies are dressed but also momenta. Dressed quasi-particles 
scatter with the scattering phase $\dr\Phi$ which differs significantly from the undressed one. As we shall see $\dr\Phi$ involves e.g. chiral-chiral interaction which is absent in the original model.
Hence its interpretation can not be as straightforward as chiral-anti-chiral processes discussed in \cite{2012JHEP...09..133D}. 

%
\paragraph{Dressed momentum} \label{sec:p-dr}
Momentum of quasi-particles are dressed according to
\begin{align}
\dr{p}_A=&p_A+\sum_B\int \sl{dq} \Phi_{AB}^{(0)} (p,q) \rho_B(q).
\end{align}
This can rewritten explicitly as
\begin{align}\label{p-dr2}
\dr{p}{}^+&=p^+\left(1-\isl{k}\av{-}\right),\quad
\dr{p}{}^-=p^-\left(1+\isl{k}\av{+}\right)
\end{align}
The rescaling factor depends on chirality only.
It changes definition of chirality for $\dr p$'s what may create strange momenta gap (or overlap): for chiral quasi-particles  $p>-\mu_A$ means $\dr p{}^+>-\mu_A (1-\isl{k}\av{-})$
while for anti-chiral quasi-particles $p<-\mu_A$ means $\dr p{}^-<-\mu_A (1+\isl{k}\av{+})$.
Momenta of the quasi-particles are not observable unless they respect BE thus the gap does not necessary signal troubles but it is convenient to have $\av{-}=-\av{+}$ what removes the gap. The conditions under which this happens will be discussed in 
Sec.\ref{sec:cp}
\paragraph{Dressed energy}
Excitations above the thermal bath are given by \eqref{e-dr}
devided by $\b$ \cite{Korepin:1993kvr}. We rewrite it as
\begin{align}\label{H-dr}
{\dr H_A}^+(q)&=~~~\k^+(q+\mu_A^+)+\be^+_A,\;\quad q>-\mu_A
\non
{\dr H_A}^-(q)&=-\k^-(q+\mu_A^-)+\be^-_A,\;\quad q<-\mu_A, 
\end{align}
where $\k^\pm\equiv 1-I^\mp/\b$.
Thus due to simplicity of interaction the dispersion relation is a renormalized version of \eqref{energy-0} with major impact of potentials \eqref{pot}. 
We let general chiral $\mu^\pm_A$ replacing $\mu_A$ due to dressing of momenta.
New  couplings $\k^\pm$ and $\be^\pm_A$ signal chiral-anti-chiral disparity if $\k^+\neq \k^-$.

\paragraph{Dressed scattering phase $\dr\Phi$.} It is given by the integral equation
\label{sec:scatt-dr}
\begin{equation}
{\dr\Phi_{AB}(p,q)+\sum_C\int\sl{dr} \p_r\Phi^{(0)}_{AC}(p,r) n_C(r)
\dr\Phi_{CB}(r,q)=\Phi^{(0)}_{AB}(p,q)}
\end{equation}
The equation can be easily solved by
\begin{align}
\label{scatt-dr}
\dr\Phi_{AB}(p,q)&= C_{++}p_A^+q_B^+ +C_{+-}p_A^+q_B^-
+C_{-+}p_A^-q_B^++C_{--}p_A^-q_B^-
\end{align}
for
\begin{align}
C_{+-}&=-C_{-+}=\frac{-\isl{k}}{1+(\isl{k})^2\av{-}\av{+} }\non
C_{++}&=\frac{\av{-}}{(\sl{k})^2+\av{-}\av{+} },\qquad
C_{--}=\frac{\av{+}}{(\sl{k})^2+\av{-}\av{+} }\nn
\end{align}
It is interesting that $\dr\Phi$ involves interaction between quasi-particles of the same 
chirality. It is clear the effect results from interaction through thermal bath. In consequence 
the corresponding  BE is more involved and  non-trivial even for single quasi-particle excitation above the bath: see Sec.\ref{sec:mirror}.

\section{The new finite size model}\label{finite}
The results of the previous section shows that in thermal bath renormalizes basic components of the initial model under consideration. In particular  scattering quasi-particles of the same chirality makes sense. We take these results as our starting point. 
Thus we set the dispersion relation to be given by \eqref{H-dr} and take the scattering phase in general form \eqref{scatt-dr}. Quasi-particles of both chiralities will be treated completely independently.

\subsection{Mirror TBA}
\label{sec:mirror}
Mirror \cite{2007JHEP...12..024A,2010JHEP...05..031A} TBA allows to calculated energies of states at finite volume $R$
\cite{Dorey_1996,2012LMaPh..99..299B}.
The mirror transformation is 
\begin{equation}\label{mirror}
 p\to i \mi H,\quad  H\to i \mi p
\end{equation}
Mirroring of  \eqref{H-dr} gives
\begin{align}\label{H-mi}
\mi H^+_A=&~~~~\frac{1}{\k^+}(\mi p+i \bep_A)+i\mu_A^+=~~~~\frac{1}{\k^+}\mi p+i \mf_A^+, \quad&&\mi p>\bep_A\\
\mi H^-_A=&-\frac{1}{\k^-}(\mi p+i \bet_A)+i\mu_A^-=-\frac{1}{\k^-}\mi p+i \mf_A^-&&\mi p<- \bet_A\nn
\end{align}
where $\mf_A^\pm=\pm\frac{\be_A^\pm}{\k^\pm}+\mu_A^\pm$.
Notice the  ranges of chiral and anti-chiral mirror momenta which follows from \eqref{H-dr} and analogous construction in \cite{2020arXiv201002782S}. This guarantee that the real value  of $\mi H$ is non-negative.


The mirror scattering phase \cite{2007JHEP...12..024A} follows from  the dressed one \eqref{scatt-dr} and 
respects $\mi\Phi(\mi p,\mi q)=\dr\Phi(p,q)$ i.e.
\beq\label{scatt-mi} 
{\mi\Phi}_{AB}(\mi p,\mi q)= \dr\Phi_{AB}(i\mi H(\mi p),i\mi H(\mi q))
\eeq
In order to find the energies of states we need to write TBA for mirror pseudo-energies.
\begin{align}
\label{TBA-mi-1}
\mi\ep_A(\mi p)= &\;\psi_A+R\, \mi H_A(\mi p)-[\La_B*\mi\phi_{BA}]-i\ \mbox{$\sum_j$} \mi\Phi_{AB_j }(\mi p,\mi p_j)\,,\\
\label{ep-p-mi}
\mi\ep_{A_j}(\mi p_j)=&-2 \pi  i\, n_j+\psi_{A_j}
\end{align}
where $\mi\phi_{BA}(\mi p,\mi q)=\ \p_{\mi p}\mi\Phi_{BA}(\mi p,\mi q)$
and
\begin{align}
[\La_B*\mi\phi_{BA}]&=\int\sl{d\mi q}\La_B(\ep(\mi q))\mi\phi_{BA}(\mi q,\mi p)\non
&=
-(C_{++}\mi H^+_A(\mi p)+C_{+-}\mi H^-_A(\mi p))\mi I^++(C_{-+}\mi H^+_A(\mi p)+C_{--}\mi H^-_A(\mi p))\mi I^-\nn
\end{align}
We have introduced two integrals (mirror counterparts of  to \eqref{Is})
\begin{align}
\label{I-mi}
\mi I^\pm=\frac1{{\k}^\pm}\sum_A\int_{\mi c^\pm_A}\frac{dp}{2\pi} \La_A(\mi\ep_A^{\,\pm}(p))
\end{align}
where  ${\mi c}^\pm_A$ are  (as dictated by \eqref{H-mi}): 
$\mi c^{\,+}_A=\{{\bep_A}, +\infty\},\
\mi c^{\,-}_A=\{-\infty, -{\bet_A}\}$.
Notice that (contrary to standard kinematics TBA \eqref{TBA-0}) due to $\psi_A$ appearing in the solution \eqref{e-mi-sol} bosonic and fermionic quasi-particles contributions differ only by on overall sign.

From \eqref{scatt-mi} and  \eqref{TBA-mi-1} we infer that mirror pseudo-energies depend on flavour  only through chirality and have the form:
\begin{equation}\label{e-mi-sol}
\mi\ep_A^\pm(\mi p)=\psi_A+g^\pm\, \mi H_A^\pm(\mi p^\pm)
\end{equation}
with  two unknown constants $g^\pm$ determined by the mirror TBA \eqref{TBA-mi-1}.
The constants have interpretation of an effective string size as we shall see soon.

Substituting \eqref{e-mi-sol} to mirror TBA we get equations on $g^\pm$ 
\begin{align}\label{mi-g}
(g^+-R-(C_{++}P^++C_{+-}P^-))g^-=&\ \frac{g^-}{g^+}C_{++}\,\mi s^+-C_{-+}\,\mi s^-,\\
(g^--R-(C_{--}P^-+C_{-+}P^+))g^+=&\ C_{+-}\,\mi s^+-\frac{g^+}{g^-}C_{--}\,\mi s^-.
\end{align}
where $P^\pm=\sum_i p^\pm_i$ and $\mi s^\pm=-\sum_Ae^{\psi_A}\int_0^\infty\sl {dp}\log(1-e^{-p-g^\pm(\be^\pm_A/\k^\pm+i\mf^\pm_A)})$. We assume that $\be_A^\pm/\k^\pm>0$ then $|\mi s^\pm|\leq \pi^2$.
All chiral chemical potentials enter through $\mi s^\pm$'s only.  Thus we can treat $\mi s^\pm$'s as free parameters.  For reality of $g$'s we need reality of $\mi s$'s. 
This puts constraints on $\be$'s.  Moreover for supersymmetric backgrounds $\be$'s (so also $\mf$'s) for bosons and fermions are the same thus $\mi s^\pm=0$ for any $g$'s thus also  for any $N$'s. 

It is worth to notice that $g$'s are sums of three terms: bare size of the string $R$,
the total chiral momenta $P^+,P^-$ 
of  the given state and finally, displayed on the r.h.s., the contributions of virtual particles  circling around the string. The latter vanish for 
$R\to \infty$ (thus also $g^\pm\to \infty$) due to $\be_A^\pm/\k^\pm>0$. 

\subsubsection{States}
\paragraph{Ground state}
In \eqref{mi-g} we set $P^\pm=0$.
The equations determine $g^\pm$ as functions of $R$ and $\be$'s. 
The  ground state energy is
\begin{equation}\label{ground}
H_0 = - \frac{1}{2\pi}(\mi{p}_A)'* \Lambda_A\,=
-(\k^+\frac{\mi s^+}{g^+}+\k^-\frac{\mi s^-}{g^-})
\end{equation}
Supersymmetric backgrounds ($\mi s^\pm=0$) yields $g^+=g^-=R$ and $H_0=0$. 

\paragraph{Excitations}
We need to consider full \eqref{mi-g}.  We can rewrite it using  condition 
\eqref{ep-p-mi} i.e.   $2 \pi  n^\pm_j=\, g^\pm p^{\pm}_j$ what is equivalent to
$2 \pi  N^\pm=\, g^\pm P^{\pm}$ with $N^\pm=\sum_j n^\pm_j$ and $ P^\pm=\sum_j p^\pm_j$.  Hence we obtain equations on $P^{\pm}$ in terms of 
integers $N^\pm$ and $\mi s^\pm$.
The total chiral momenta determine quasi-particles momenta $p^\pm_i$ in terms od $n^\pm_i$. At that point we need to check if the assignment of chiralities was proper i.e. in agreement with definition of chirality for given set of flavours.

At large $R$ the equations simplify to set of simple algebraic equations due $\mi s^\pm\approx 0$ (for $\be^\pm/\k^\pm\geq0$).
\begin{align}\label{BE-dr}
p_{i}^+(R+C_{++}P^++C_{+-}P^-)&=2\pi n_{i}^+,\\
p_{i}^-(R+C_{-+}P^++C_{--}P^-)&=2\pi n_{i}^-
\end{align}
The multi-quasi-particle case
 can be solved almost as easily as the original model with $C_{--}=C_{++}=0$.
Notice that the above equations are non-trivial even for single quasi-particle excitation, e.g.
\begin{equation}
\label{BE-dr-1}
p^+(R+C_{++}p^+)=2\pi n^+
\end{equation}

Energy of excitations is given by (\cite{Dei_2018} Eq.(3.19))
\begin{align}
\label{TBA-mi-energy}
H_{tot}=&- \frac{1}{2\pi}(\mi{p}_A)'* \Lambda_A+\sum_j \dr E(p_j)\\
=&
-(\k^+\frac{\mi s^+}{g^+}+\k^-\frac{\mi s^-}{g^-})
+\k^+P^+-\k^-P^-+\sum_{j^+}(\k^+\mu_{A_j}+\be^+_{A_j})
+\sum_{j^-}(-\k^-\mu_{A_j}+\be^-_{A_j}).\nn
\end{align}
where $j^\pm$ enumerates chiral $(+)$ or anti-chiral $(-)$ quasi-particles. 

\subsection{CP invariance}\label{sec:cp}

In general the deformed model under considerations contain a lot of new parameters: 
e.g. $\be$'s which are not observable and in principle can even be complex although in  this paper they are assumed to be real.
The space of free parameters can be simplified imposing some natural  discrete symmetries. 
We can define cousins of well known C,P symmetries of the ordinary relativistic QFT. 
Notice that \eqref{energy-0}
is invariant under $p\to-p$ and $A\to\bar A$ i.e.
\beq
H^{(0)}_A(p)=H^{(0)}_{\bar A}(-p)
\eeq
The two combined transformations we shall call CP. 
For thermodynamical  quantities defined in Sec.\ref{sec:TBA} CP invariance requires
$\be^+_A=\be^-_{\bar A},\ \mu^+_A=-\mu^-_{\bar A}$ and $\k^+=\k^-\equiv \k$ .
Then we have $\cI(\be^+)=\cI(\be^-)$. It follows that $I^+=I^-\equiv I$ is the solution to  \eqref{Is-eq} 
which goes to zero when $\b\to\infty$ i.e. $T\to 0$.  Also
$\av{+}=-\av{-}$ what will be important for the scattering phase invariance.
Formulae of Sec.\ref{sec:scatt-dr} express $C_{\pm\pm}$ in terms of $\av{\pm}$.
This gives
 $C_{++}=-C_{--},\ C_{+-}=-C_{-+}$.
For the mirror  quantities we get
$(\mi H_{\bar A}(-p))^*=\mi H_A(p)$, 
$\mf_A^+=-\mf_{\bar A}^-$, so $\mi s^+\oto\mi s^-$ for $g^+\oto g^-$.


\subsubsection{CP invariance of the spectrum}
We shall show that CP-invariant Hamiltonian and the scattering phase as discussed above yields by mirror BE \eqref{mi-g} CP-invariant spectrum i.e. to any state characterized by two sets of quantum numbers
$( \qn^+=\{(n_i^+,A_i)\}, \qn^-=\{n_j^-,A_j\})$  there is a CP-conjugate state characterized by 
$({\qn'}^+=\{-n_j^-,\bar A_j\}, {\qn'}^-=\{-n_i^+,\bar A_i\})$ of the  same energy.

First of all we check good quantum numbers assignment: indeed  chirality of  $(n_i^+,A_i)$ i.e.  $p_i^+>-\mu_A$ implies anti-chirality of $(-n_i^+,\bar A_i)$ i.e. $-p_i^+<-\mu_{\bar A_i}$. Next we show that both $(\qn^+,\qn^-)$ and $ ({\qn'}^+,{\qn'}^-)$ respect the  CP invariant version of  \eqref{mi-g} which is:
\begin{align}\label{mi-g-cp}
(g^+-R-(C_{++}P^++C_{+-}P^-))g^-=&\ \frac{g^-}{g^+}C_{++}\,\mi s^++C_{+-}\,\mi s^-,\\
(g^--R-(-C_{++}P^--C_{+-}P^+))g^+=&\ C_{+-}\,\mi s^++\frac{g^+}{g^-}C_{++}\,\mi s^-.
\end{align}
Recall $g^\pm=2\pi N^\pm/P^\pm=n_j^\pm/p_j^\pm$ (for all $j$).
 It is enough to notice that \eqref{mi-g-cp} is invariant under 
$n_j^+\oto -n_j^-$ and  $P^+\oto -P^-$ (what implies 
$g^+\oto g^-$ and $\mi s^+\oto \mi s^-$).
Finally,   one can easily check invariance of the energy formula.

\section{Conclusions}
We have discussed massless integrable model with scattering phase given by product of quasi-particles momenta.  As starting point we took \adsst string model of \cite{Dei_2018} for which only chiral-anti-chiral quasi-particles interact directly.
Analyzing thermodynamics we have shown that the effective (dressed) scattering phases involve also chiral-chiral and anti-chiral-anti-chiral terms. 
This suggests that models involving direct interactions of all chiralities are natural,
despite possible problems with  interpretation of such scattering. On the other hand one must remember that we always work with finite size circular space for which naive flat  space intuitions may not work.
Although the discussed models 
can be considered as deformations of the original \adsst string, they are  interesting on its own. 

We have also modified spectral relation including general chemical potentials. We have derived mirror TBA equations which provide equation for finite  size spectrum. It is interesting that one can impose string counterparts of well known CP symmetry
which  (if supplemented by T transformation) is necessary ingredients of  any relativistic invariant theory. Hence the symmetry might be required for any consistent integrable model of the type discussed here if it corresponds to a string theory compactification. 
At first step these issues could be discussed on the level of small deformations of \adsst strings models.  We leave this problem to the future publication.

\section*{Acknowledgments}
I would like to thank Zoltan Bajnok, Roman Janik, and Mi{\l}osz Panfil  for insightful discussions about integrable models as well as   Andrea Cavaglia,
Nikolay Gromov,
Bogdan Stefanski and
Alessandro Torrielli for interesting email exchange.
\bibliographystyle{../tex/JHEP}
\bibliography{../tex/refs}
\end{document}